\documentclass{emulateapj}
\shorttitle{Bulk comptonization at supernova shock breakout}
\shortauthors{SUZUKI\&SHIGEYAMA}
\begin{document}
\title{NON-THERMAL PHOTON PRODUCTION VIA BULK COMPTONIZATION AT SUPERNOVA SHOCK BREAKOUT}
\author{AKIHIRO SUZUKI\altaffilmark{1,2} and TOSHIKAZU SHIGEYAMA\altaffilmark{1}}
\altaffiltext{1}{Research Center for the Early Universe, School of Science, University of Tokyo, Bunkyo-ku, Tokyo 113-0033, Japan.}
\altaffiltext{2}{Department of Astronomy, School of Science, University of Tokyo, Bunkyo-ku, Tokyo 113-0033, Japan.}
\begin{abstract}
We investigate roles of the bulk comptonization process in the propagation of thermal photons emitted at the shock breakout of a supernova explosion. 
We use a hydrodynamical model based on a self-similar solution for the shock breakout. 
The  propagation of photons is treated by a Monte-Carlo technique. 
Results of the simulations successfully reproduce the power-law spectrum observed by X-ray observations for XRF 080109/SN 2008D, if a mildly relativistic shock velocity is assumed. 
Effects of some radiative processes, photoionization, radiative recombination, and free-free absorption on the propagation of emitted photons are also investigated. 
It is found that none of them hardly changes the spectrum regarding the progenitor stars of type Ib or Ic supernovae. 
Light curves  calculated under the assumption of a spherical explosion indicate that the progenitor radius is required to be $\sim 10^{13}$ cm. 
\end{abstract}

\section{INTRODUCTION\label{intro}}
The collapse of the iron core via photo-disintegration of nuclei at the final evolutionary phase of a massive star results in the formation of a strong shock wave in the deep interior. 
The shock wave propagates in the envelope, finally reaches the surface, and outshines the emission from the progenitor. 
This phenomenon occurring at the beginning of a supernova explosion is called supernova shock breakout\citep[][]{kc78,f78,mm99}. 
A growing example of detections of the shock breakouts, e.g., SN 2008D \citep[][]{s08,l08,m08,m09}, SNLS-04D2dc \citep[][]{sc08,g08}, SNLS-06D1jd \citep{g08}, promotes investigations into its property, detectability, and usage for cosmological studies \citep[e.g.][]{to09}. 

Among the example detections, the X-ray outburst XRF 080109 was serendipitously detected during the follow-up observations of a previously discovered Type Ibc supernova, SN 2007uy \citep{s08}. It was associated with another Type Ib/c supernova 2008D. The parent galaxy was a spiral galaxy NGC 2770 at the distance of 28 Mpc. Analyses of the observed X-ray spectrum and light curve have evoked a controversy on the origin of the emission. 
Some authors attribute the origin of the X-ray emission to the emergence of a radiation-dominated shock wave from the envelope of the progenitor, i.e., the supernova shock breakout. 
Others consider that SN 2008D is a gamma-ray burst (GRB) associated supernova and the X-ray emission is originated from the ejected matter moving at relativistic speeds. 

\cite{s08} presented the X-ray light curve of XRF 080109 and results of the spectral analysis.
The emission lasted for $\Delta t \sim 600$ s and its peak luminosity was $L_\mathrm{X}=6.1\times 10^{43} \,\mathrm{erg}\ \mathrm{s}^{-1}$. 
The total energy of the outburst is $E_\mathrm{X}=2\times 10^{46}\,\mathrm{erg}$. 
The X-ray spectrum is best fitted by a single power-law with the photon index of $\Gamma=2.3\pm0.3$. 
Assuming the presence of relativistic ejecta, they estimated the bulk Lorentz factor at $\gamma\sim 90$. 
Because the scale length of the emission region, $\gamma^2c\Delta t\sim 10^{17}$ cm, is much larger than the expected radius of the progenitor star, $R=10^{10}$ cm, they excluded the possibility of the relativistic ejecta origin of X-ray photons. Here $c$ denotes the speed of light. \cite{l08} performed a spectral analysis of XRF 080109 and found that the X-ray spectrum is well fitted by a power law spectrum with the photon index of $\Gamma=2.29^{+0.28}_{-0.26}$  absorbed by matter with the total equivalent hydrogen column density of $N_H=6.83^{+1.5}_{-1.3}\times 10^{21}\,\mathrm{cm}^{-2}$, leading to a conclusion that the shock breakout origin is unlikely because no black body component is found. 
\cite{m08} also insisted on the relativistic ejecta origin because of little contribution from a black body component to the X-ray spectrum. 
However, \cite{m09} claimed that the X-ray spectrum is well fitted by a black body combined with a power law spectrum with 50\% bolometric contribution from the black body component. 
Their best fit values for the photon index $\Gamma$ and the black body temperature $T_\mathrm{BB}$ are $\Gamma=2.1^{+0.3}_{-0.4}$ and $kT_\mathrm{BB}=0.10\pm0.01$ keV, respectively. 

While X-ray observations have not led to agreement on the origin of the X-ray emission, observations in other wavelength ranges support the shock breakout origin. 
\cite{b09} have performed very-long-baseline interferometry (VLBI) observations at 30 and 133 days after the explosion. 
They succeeded in resolving the sizes of the source at both epochs and determined the $3\sigma$ upper limit to the expansion velocity to be $\sim 0.75c$ , which rules out the presence of ultra-relativistic ejecta.
From optical spectroscopic observations for SN 2008D in the nebular phase, \cite{t09} found a double-peaked [\ion{O}{1}] line profile, which is strong evidence of an aspherical explosion. 
They argued that SN 2008D is a side-viewed bipolar explosion.

If relativistic ejecta are responsible for the observed X-ray emission, X-ray photons will be emitted by synchrotron mechanism, which naturally explains the observed power-law (or black body + power law) X-ray spectrum. 
On the other hand, early investigations into the supernova shock breakout \citep[][]{kc78,f78,mm99} predicted thermal black body radiation for a UV/X-ray flash associated with the shock breakout.  Therefore, some additional processes producing non-thermal photons are required to explain the observed power-law spectrum in the framework of the shock breakout origin. 
The bulk comptonization is a process to produce non-thermal photons in the presence of a bulk flow of matter \citep{bp81a,bp81b}.
A small fraction of photons are repeatedly scattered by electrons across the shock front and gradually increasing their energies. 
The first application of the process to the supernova shock breakout was done by \cite{wwm07}, who calculated the propagation of photons around  the front of a radiation-dominated shock by a Monte-Carlo  method. 
However, the shock profiles assumed in their calculations might oversimplify the situation and effects of other radiative processes competing the electron scattering, e.g., photoionization, radiative recombination, and free-free absorption, were not discussed. In addition, they did not construct a light curve from their model.
Hence we try to establish a more realistic model than \cite{wwm07} by using an appropriate hydrodynamical model, the self-similar solution for the supernova shock breakout. 
The solution originally derived by \cite{s60} describes the propagation of a shock wave in a planar atmosphere with equations of non-relativistic fluid dynamics. 
We treat the radiative transfer of thermal photons emitted by the shock front in the test-particle limit. 
Then we construct not only spectra but also X-ray light curves from models with several different parameter sets. 
To derive the spectra, we neglect the contribution from photons inside the photosphere, which is defined as the position where the optical depth measured from the stellar surface becomes unity. 
Using the results, we discuss the origin of the X-ray emission from XRF 080109.

This paper is structured as follows. 
In \S 2, we describe formulation of the problem. 
The derivation of the self-similar solution for the shock breakout used in \S 2 is reviewed in Appendix. 
An application of our model to XRF 080109/SN 2008D is presented in \S 3. 
Finally, \S 4 concludes this paper.  

\section{FORMULATION}
In this section, we explain qualitative natures of the supernova shock breakout, describe details of our model, and formulate the problem. 

The collapse of the iron core of a massive star results in the formation of a strong shock in the deep interior and the shock propagates toward the envelope. 
When propagating in the envelope, the shock is mediated by radiation. 
In the diffusion approximation, the average velocity $v_\mathrm{diff}$ of photons diffusing in a medium with the optical depth $\tau$ is expressed by the speed of light divided by the optical depth, $v_\mathrm{diff}=c/\tau$. 
Here $\tau$ represents the optical depth from the outer edge of the envelope to the shock front. 
When the shock velocity $V_\mathrm{s}$ is larger than the diffusion velocity, photons behind the shock front are trapped . 
On the other hand, when the shock velocity is smaller than the diffusion velocity, photons can overtake the shock front and observers can see these photons as a precursor of the shock front. 
Therefore, when the relation $V_\mathrm{s}=v_\mathrm{diff}$ is satisfied, namely, the optical depth decreases to the ratio of the speed of light to the shock velocity, $\tau=c/V_\mathrm{s}$, the shock breakout occurs. 

\subsection{Shock profiles}
A semi-analytical description of time evolutions of hydrodynamical variables at the supernova shock breakout has been obtained by \cite{s60}. 
The solution was extended to the ultrarelativistic case by \cite{ns05}. 
Furthermore, the present authors extended the solution so that it can treat the energy deposition or loss at the shock front \citep{ss07}. 
In this work, we use the self-similar solution discovered by \cite{s60} to obtain velocity and density profiles behind the shock front. 
In the following subsections, we review the approach of \cite{s60} and describe the procedure to obtain the profiles. 

\subsubsection{Self-similar solution}
\cite{s60} studied the propagation of a strong shock in a medium whose density profile $\rho_0(x)$ can be written by a power of the distance $x$ measured from the outer edge of the envelope of the progenitor star, 
\begin{equation}
\rho_0(x)=\left\{\begin{array}{lll}k_1x^\delta&\mathrm{for}&x\geq0,\\
0&\mathrm{for}&x<0
\end{array}\right.
\end{equation}
where $k_1$ and $\delta$ are constants characterizing the density structure of the envelope.  
Because we are concerned with the radiative envelope of a massive star, we take $\delta=3$ as the fiducial value. 
The value of $k_1$ is treated later. 
Time evolutions of the velocity $u(x,t)$, the density $\rho(x,t)$, and the pressure $p(x,t)$ behind the shock front are governed by the following equations for adiabatic gases in the plane parallel atmosphere, 
\begin{eqnarray}
\frac{\partial \rho}{\partial t}+\frac{\partial(\rho u)}{\partial x}&=&0,\label{mass}\\
\frac{\partial u}{\partial t}+u\frac{\partial u}{\partial x}+\frac{1}{\rho}\frac{\partial p}{\partial x}&=&0,\\
\frac{\partial }{\partial t}\left(\frac{p}{\rho^\gamma}\right)+u\frac{\partial }{\partial x}\left(\frac{p}{\rho^\gamma}\right)&=&0\label{entropy},
\end{eqnarray}
where $\gamma$ represents the adiabatic index. 
In this paper, $\gamma$ is fixed to $4/3$ because the shock is assumed to be radiation-dominated. 
The time $t$ is measured from the point when the shock reaches the outer edge of the envelope, $x=0$.  
For the self-similarity, we assume that the shock velocity $V_\mathrm{s}$ is expressed by a power of the position $X$ of the shock, 
\begin{equation}
V_\mathrm{s}=k_2X^{-\lambda},
\label{Vs}
\end{equation}
where $k_2$ is a constant and we have introduced a parameter $\lambda$, the value of which is determined so as to satisfy appropriate boundary conditions. 
The procedure to determine $\lambda$ is summarized in Appendix. 
Integrating the definition of the velocity of the shock, $dX/dt=V_\mathrm{s}$, with respect to time, the position of the shock is obtained as a function of time $t$,
\begin{equation}
X=[k_2(1+\lambda)(-t)]^{1/(1+\lambda)}.
\end{equation}
Then we define the similarity variable $\xi$ as
\begin{equation}
\xi=\left(\frac{X}{x}\right)^{\lambda+1},
\end{equation}
and introduce the non-dimensional velocity $f(\xi)$, pressure $g(\xi)$, and density $h(\xi)$ as 
\begin{eqnarray}
u(x,t)&=&k_2X^{-\lambda}f(\xi),\nonumber\\
p(x,t)&=&k_1k_2^2X^{\delta -2\lambda}g(\xi),\\
\rho(x,t)&=&k_1X^\delta h(\xi).\nonumber
\label{relation}
\end{eqnarray}
Using these functions, the governing equations (\ref{mass})-(\ref{entropy}) are converted to ordinary differential equations,
\begin{eqnarray}
(1-\xi f)\frac{d\ln h}{d\xi}-\xi\frac{d f}{d\xi}&=&-\frac{\delta -\lambda}{1+\lambda},\label{dfdx}\\
(1-\xi f)\frac{df}{d\xi}-\frac{\xi}{h}\frac{dg}{d\xi}&=&\frac{\lambda}{1+\lambda}f^2-\frac{\delta-2\lambda}{1+\lambda}\frac{g}{h},\label{dgdx}\\
(1-\xi f)\frac{d\ln g}{d\xi}-\gamma\xi\frac{df}{d\xi}&=&-\frac{\delta-(\gamma+2)\lambda}{1+\lambda}f.\label{dhdx}
\end{eqnarray}

Next, we consider the boundary condition at the shock front. 
The Rankine-Hugoniot relations for a strong shock determine the values of the velocity $u_\mathrm{f}$, the pressure $p_\mathrm{f}$, and the density $\rho_\mathrm{f}$ immediately behind the shock front,
\begin{equation}
u_\mathrm{f}=\frac{2}{\gamma+1}V_\mathrm{s},\ \ \ 
p_\mathrm{f}=\frac{2}{\gamma+1}\rho_0V_\mathrm{s}^2,\ \ \ 
\rho_\mathrm{f}=\frac{\gamma+1}{\gamma-1}\rho_0.
\end{equation}
The boundary condition at the shock front yields
\begin{equation}
f(1)=\frac{1}{\gamma+1},\ \ \ g(1)=\frac{2}{\gamma+1},\ \ \ h(1)=\frac{\gamma+1}{\gamma-1}.
\label{bc}
\end{equation}
 Integrating Eqs. (\ref{dfdx})-(\ref{dhdx}) from $\xi=1$ to $\xi =0$ for the eigen value of $\lambda$, we obtain the profiles of $f(\xi)$, $g(\xi)$, and $h(\xi)$. Otherwise the solution could not reach $\xi=0$.

\subsubsection{Derivation of the shock profiles}
Once we obtain the profiles, $f(\xi)$, $g(\xi)$, and $h(\xi)$, we can derive the time evolutions of hydrodynamical variables $u(x,t)$, $\rho(x,t)$, and $p(x,t)$ by using the relations (\ref{relation}). 
To apply the solution to a specific problem, we need to fix the free parameters $k_1$ and $k_2$ introduced above. 

To fix them, we determine the position $X_\mathrm{i}$ and velocity $V_\mathrm{i}$ of the shock front at the shock breakout when the optical depth is equal to the ratio of the speed of light to the shock velocity. 
Since we can calculate the optical depth $\tau$ by simply integrating the density profile ahead of the shock front, 
\begin{equation}
\tau=\frac{\sigma_\mathrm{T}}{\mu m_\mathrm{H}}\int _0^{X_\mathrm{i}}k_1x^\delta dx
=\frac{\sigma_\mathrm{T}k_1X_\mathrm{i}^{\delta +1}}{(\delta +1)\mu m_\mathrm{H}},
\label{eq14}
\end{equation}
where $\mu(=0.6)$ is the mean molecular weight, $m_\mathrm{H}$ the Hydrogen mass, and $\sigma_\mathrm{T}$ is the Thomson cross section, we obtain the value of $k_1$ for given $X_\mathrm{i}$ and $V_\mathrm{i}$ as,
\begin{equation}
k_1=\frac{(\delta+1)\mu m_\mathrm{H}c}{\sigma_\mathrm{T}X_\mathrm{i}^{\delta +1}|V_\mathrm{i}|}.
\label{eq15}
\end{equation}
The parameter $k_2$ is obtained from the definition of the shock velocity (\ref{Vs}),
\begin{equation}
k_2=V_\mathrm{i}X_\mathrm{i}^\lambda.
\end{equation}
Next, we define a value $X_\mathrm{f}$ so that the optical depth $\tau$ becomes unity when the shock reaches the position $x=X_\mathrm{f}$. 
Using Eqs. (\ref{eq14}) and (\ref{eq15}), the position $X_\mathrm{f}$ is expressed as 
\begin{equation}
X_\mathrm{f}=\left(\frac{|V_\mathrm{i}|}{c}\right)^{1/(1+\delta)}X_\mathrm{i},
\end{equation}
which corresponds to the final velocity $V_\mathrm{f}$ of the shock wave given by
\begin{equation}
V_\mathrm{f}=\left(\frac{|V_\mathrm{i}|}{c}\right)^{-\lambda/(1+\delta)}V_\mathrm{i}.
\end{equation}
We stop the calculation for the photon propagation when $X=X_\mathrm{f}$. 

\begin{figure}
\begin{center}
\includegraphics[scale=0.5]{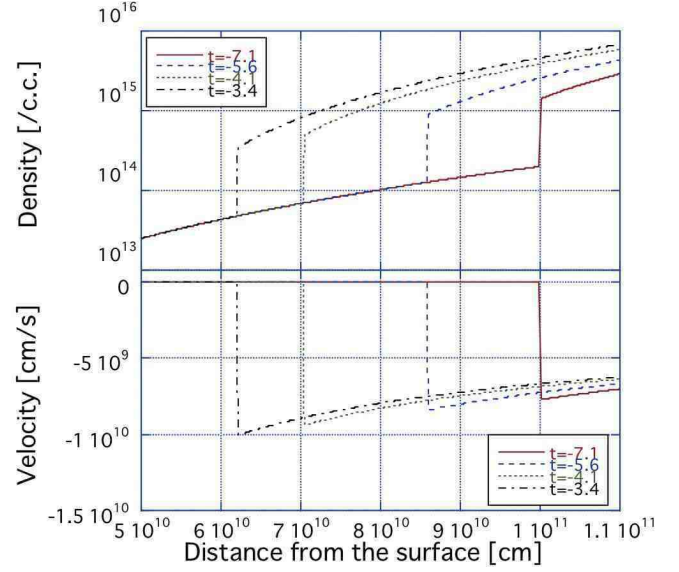}
\caption{Snapshots of the density profile (upper panel) and the velocity profile (lower panel). 
Each line represents the density or the velocity at $t=-7.1$ s (solid), $-5.6$ s (dashed), $-4.1$ s (dotted), and $-3.4$ s (dash-dotted). }
\label{f1}
\end{center}
\end{figure}

An example of the thus derived profiles of the velocity $u(x,t)$ and density $\rho(x,t)$ are shown in Figure \ref{f1}. 
The free parameters are chosen as $X_\mathrm{i}=10^{11}$ cm and $V_\mathrm{i}=0.3c$.

\subsubsection{Uncertainty of the profiles}
Although the set of the derived profiles is an exact solution of the governing equations for non-relativistic adiabatic gases, it partly deviates from a realistic model for a radiation-dominated shock propagating in the stellar envelope. 
Especially, the temperature profile after the shock breakout must deviate from that derived from the density and/or pressure profiles by making use of an appropriate equation of state because the present method does not take into account the energy transfer by photons overtaking the shock front. 
Therefore, in the following calculations of the photon propagation around the shock front, we only use two profiles, the velocity and density profiles. 
The former is used to determine the bulk velocity of electrons scattering photons and the latter is used for the calculation of the optical depth. 
On the other hand, the pressure profile indicates a roughly constant post-shock pressure $p_\mathrm{s}\simeq 10^{10}\,\mathrm{g/cm\ s}^2$.   
Since the post-shock pressure is expected to be dominated by radiation pressure, $p_\mathrm{s}=aT_\mathrm{ph}^4/3$, where $a$ is the radiation constant, we obtain the photon temperature of the order of $kT_\mathrm{ph}\simeq 0.1$ keV, which is consistent with results of spectral analysis of XRF 080109 \citep{m09}. 
Then, we assume a uniform temperature profile with $kT_\mathrm{ph}=0.1$ keV behind the shock. 


\subsection{Radiative processes\label{radiative_process}}
Before we describe the procedure to deal with the photon propagation, we estimate relative contributions of some radiative processes important in the stellar envelope. 
For photoionization and radiative recombination, we estimate the time scale of each process for oxygen, which is one of the most abundant heavy elements. 
For radiative recombination, we compare the absorption coefficient with the scattering coefficient for Compton scattering. 
In the following discussion, we assume the envelope of a compact star such as Wolf-Rayet stars because the velocity of the shock wave propagating in the envelope is expected to reach mildly relativistic speeds, which is required for the bulk comptonization to work well.

First of all, we estimate the average electron number density $\bar{n}_\mathrm{e}$ of the envelope when the electron scattering is a dominant source of opacity. 
Then, we suppose that the shock front is located at a distance $\Delta R$ from the outer edge of the envelope. 
Using the average number density $\bar{n}_\mathrm{e}$, we can express the optical depth from the outer edge of the envelope to the shock front as $\bar{n}_\mathrm{e}\sigma_\mathrm{T}\Delta R$. 
Since the optical depth at the moment of the shock breakout is given by $\tau=c/V_\mathrm{s}$, the average electron number density of the envelope is estimated as, 
\begin{equation}
\bar{n}_\mathrm{e}=\frac{c}{\sigma_\mathrm{T}\Delta RV_\mathrm{s}}=
3\times 10^{15}\left(\frac{\Delta R}{10^{9}\,\mathrm{cm}}\right)^{-1}
\left(\frac{V_\mathrm{s}}{0.3c}\right)^{-1}\,\mathrm{cm}^{-3}.
\end{equation}
Therefore, for a progenitor star with the radius $R$, the number of electrons ahead of the shock front is estimated as
\begin{eqnarray}
4\pi\bar{n}_\mathrm{e}\Delta R R^2&=&4\times 10^{47}\left(\frac{\bar{n}_\mathrm{e}}{3\times 10^{15}\,\mathrm{cm}^{-1}}\right)\nonumber\\
&&\ \ \ \ \ \times \left(\frac{\Delta R}{10^{9}\,\mathrm{cm}}\right)\left(\frac{R}{10^{11}\,\mathrm{cm}}\right)^2.
\label{Nph}
\end{eqnarray}

\subsubsection{Photoionization\label{photoionization}}
Next, we consider photoionization of heavy elements. 
X-ray photons emitted from the shock front ionize elements in the envelope. 
Especially, heavy elements like oxygen and neon efficiently absorb the X-ray photons because the absorption edges of such elements are in the energy range of the X-ray photons. 
For example, the photoionization cross section of Hydrogen-like oxygen ion for photons with energy $h\nu$ is given by
\begin{equation}
\sigma_\mathrm{ph}=\left(\frac{\pi g}{3\sqrt{3}}\right)\alpha_\mathrm{fs} a_0^2\left(\frac{\chi}{h\nu}\right)^3\ \ \ \mathrm{for}\ h\nu>\chi,
\end{equation}
\citep{rl85}, 
where $g(\sim 1)$ is the gaunt factor, $\alpha_\mathrm{fs}$ is the fine structure constant, and $a_0$ is the Bohr radius. 
The energy $\chi$ is the ionization potential for oxygen, 
\begin{equation}
\chi=32\alpha^2m_\mathrm{e}c^2=0.87\,\mathrm{keV}.
\end{equation}
Therefore, for photons with the energy $\chi$, the photoionization cross section is of the order of $\sigma_\mathrm{ph}\simeq 10^{-19}\,\mathrm{cm}^2$, which is much larger than the Thomson cross section $\sigma_\mathrm{T}$. 
Namely, photoionization is a dominant opacity source until all the heavy elements become fully ionized. 
In fact, however, photoionization hardly deforms the X-ray spectrum because all the relevant elements become fully ionized immediately. 
To show this, we estimate the total emitted energy $E_\mathrm{X}$ of non-thermal X-ray photons. 
Assuming the photon temperature of $T_\mathrm{ph}\simeq 0.1$ keV and the emitting radius of $R=10^{11}$ cm, the luminosity $L_\mathrm{th}$ of the thermal emission is obtained by the following formula,
\begin{equation}
L_\mathrm{th}=4\pi R^2\sigma_\mathrm{SB} T_\mathrm{ph}^4= 10^{43}
\left(\frac{R}{10^{11}\,\mathrm{cm}}\right)^2\left(\frac{T_\mathrm{ph}}{0.1\,\mathrm{keV}}\right)^4\, \mathrm{erg/s},
\end{equation}
where $\sigma_\mathrm{SB}$ is the Stefan-Boltzmann constant. 
Since the time required for the shock wave to propagate across a shell with the width $\Delta R$ is given by $\Delta R/V_\mathrm{s}$, the total energy $E_\mathrm{th}$ of the thermal emission is 
\begin{eqnarray}
E_\mathrm{th}=L_\mathrm{th}\frac{\Delta R}{V_\mathrm{s}}=10^{42}&&
\left(\frac{R}{10^{11}\,\mathrm{cm}}\right)^2\left(\frac{T_\mathrm{ph}}{0.1\,\mathrm{keV}}\right)^4\nonumber\\
&&\hspace{-1em}\times\left(\frac{\Delta R}{10^9\,\mathrm{cm}}\right)\left(\frac{V_\mathrm{s}}{0.3c}\right)^{-1}\ \mathrm{erg}.
\end{eqnarray}
Furthermore, introducing a parameter $\epsilon$, we express the total emitted energy of non-thermal X-ray photons as $E_\mathrm{X}=\epsilon E_\mathrm{th}\simeq 10^{42}\epsilon\ \mathrm{erg}$. 
Once the outburst energy $E_\mathrm{X}$ is given, we can estimate the number of X-ray photons having the energy of a few keV and the number is of the order of $10^{50}\epsilon$ photons, which is larger than the number of electrons estimated above (\ref{Nph}) when a substantial contribution of non-thermal photons, $\epsilon>0.1$, is assumed. 
From the estimated value of the number of photons and the volume of the envelope, the number density yields
\begin{eqnarray}
\hspace{-1.5em}n_\mathrm{ph}&=&\frac{10^{50}\epsilon}{4\pi R^2\Delta R}\nonumber\\
&=&8\times 10^{16}\left(\frac{\epsilon}{0.1}\right)
\left(\frac{T_\mathrm{ph}}{0.1\,\mathrm{keV}}\right)^{4}
\left(\frac{\Delta V_\mathrm{s}}{0.3c}\right)^{-1}\ \mathrm{cm}^{-3},
\end{eqnarray}
Using these values, we can evaluate the time scale for photoionization as
\begin{equation}
\tau_\mathrm{ph}=\frac{1}{c\sigma_\mathrm{ph}n_\mathrm{ph}}=4\times 10^{-9}\,\mathrm{s},
\end{equation}
which is much shorter than the time scale for the shock wave to propagate in the envelope.

\subsubsection{Radiative recombination}
Then, we estimate the time scale of radiative recombination for fully ionized oxygen. 
The time scale of radiative recombination for ions with the atomic number $Z$ is expressed by
\begin{equation}
\tau_\mathrm{rad}=\frac{1}{\alpha^\mathrm{rad}_{Z}\bar{n}_\mathrm{e}}.
\end{equation}
The radiative recombination coefficient $\alpha^\mathrm{rad}_Z$ is given by
\begin{eqnarray}
\alpha^\mathrm{rad}_Z&=&5.2\times 10^{-14}\beta^{-1/2}\nonumber\\
&&\times(0.4288+0.5\ln\beta+0.469\beta^{-1/3})\ \mathrm{cm}^3/\mathrm{s},
\end{eqnarray}
\citep{s59}. 
Here the parameter $\beta$ is given by,
\begin{equation}
\beta=\frac{\chi}{kT}=8.7\left(\frac{kT}{0.1\,\mathrm{keV}}\right)^{-1}.
\end{equation}
Using the above relations and the temperature of $kT=0.1\,\mathrm{keV}$ inferred from the spectral analysis of the X-ray spectrum \citep{m09}, we obtain
\begin{equation}
\tau_\mathrm{rad}=3\times 10^{-3}Y_\mathrm{oxygen}^{-1}\left(\frac{\bar{n}_\mathrm{e}}{3\times 10^{15}}\right)^{-1}\,\mathrm{s},
\end{equation}
where $Y_\mathrm{oxygen}$ is the number of oxygen atoms per nucleon. 
Since the estimated time scale for radiative recombination is much longer than that of photoionization even for pure-O atmosphere ($Y_\mathrm{oxygen}=1$), we can safely assume fully ionized matter in the envelope.

\cite{b06} provides results of the calculation of dielectric recombination rate coefficients for hydrogen-like ions. 
His results show that the dielectric recombination rate coefficient for the fully ionized oxygen is smaller than the radiative one for the temperature range of our interest. 
Therefore, we can neglect the contribution of dielectric recombination. 

\subsubsection{Free-free absorption}
Finally, we consider free-free absorption. 
The free-free absorption coefficient for ions with the charge $Z$ is given by
\begin{equation}
\alpha _\nu^\mathrm{ff}=3.7\times 10^{8}T^{-1/2}Z^2n_\mathrm{e}n_\mathrm{i}\nu^{-3}(1-e^{-h\nu/kT})\bar{g}_\mathrm{ff}\ \mathrm{cm}^{-1},
\end{equation}
\citep[e.g.][]{rl85},
where $n_\mathrm{i}$ is the number densities of ions, $\nu$ is the frequency of the incident photon, and $\bar{g}_\mathrm{ff}$ represents the gaunt factor. 
In the following, we consider the contribution from oxygens. 
For the temperature derived by the spectral analysis of XRF 080109, $kT=0.1\,\mathrm{keV}$ \citep{m08,m09}, the free-free absorption coefficient is estimated to be $\alpha_\nu^\mathrm{ff}\simeq 10^{-13}\,\mathrm{cm}^{-1}$ for $h\nu=0.1\ \mathrm{keV}$, 
which is much smaller than the scattering coefficient for Compton scattering, $\bar{n}_\mathrm{e}\sigma_\mathrm{T}\simeq 10^{-11}\,\mathrm{cm}^{-1}$. 
Here we have assumed that the gas is fully ionized and that the gaunt factor is unity. 
On the other hand, for the surface temperature of a massive star, $T\simeq 10^4$ K, the free-free absorption coefficient for low energy photons is comparable to the scattering coefficient, $\alpha_\nu^\mathrm{ff}\simeq 10^{-10}\,\mathrm{cm}^{-1}$ for $h\nu=10\ \mathrm{eV}$. 
However, since the effective optical thickness $\tau_\ast$ of the envelope, which is given by
\begin{equation}
\tau_\ast=\sqrt{\alpha_\nu^\mathrm{ff}(\alpha_\nu^\mathrm{ff}+\bar{n}_\mathrm{e}\sigma_\mathrm{T})}\Delta R,
\end{equation}
 \citep[e.g.][]{rl85} is smaller than unity, $\tau_\ast\simeq 0.3$, even for $T\simeq 10^4$ K and $h\nu=10$ eV, most photons must escape from the envelope without being absorbed. 
As discussed in \S \ref{photoionization}, matter in front of the shock front is exposed by the X-ray photons and free electrons thereof must have higher temperatures. 
Hence we may overestimate the effective optical thickness. 
For the progenitor stars of type Ib or Ic SNe, the contribution from other elements, e.g., helium and carbon, may be dominant. 
However, the difference of the composition of the envelope, does not alter the conclusion, $\tau_\ast<1$. 
Thus, we neglect effects of free-free absorption in the following calculations. 

\subsection{Radiative transfer}
In this subsection, we describe our method for computing the generation and the propagation of photons in given velocity and density profiles. 

\subsubsection{Photon generation}
First of all, we generate seed photons having a black body spectrum with the photon temperature $T_\mathrm{ph}$ assumed to be $kT_\mathrm{ph}=0.1\,\mathrm{keV}$. 
The frequency $\nu$ of each photon is determined by using random numbers so that the distribution $f_\gamma(\nu)$ of photons obeys the following form,
\begin{equation}
f_\gamma(\nu)\propto \frac{\nu^2}{\exp(h\nu/kT_\mathrm{ph})-1}.
\end{equation} 
The direction vector of each photon is specified by the inclination angle $\theta$ and the azimuth angle $\phi$ as $(\sin\theta\cos\phi,\sin\theta\sin\phi,\cos\theta)$. 
For seed photons, the angles $\theta$ and $\phi$ are determined so that the angular distribution of photons becomes isotropic. 
In our model, the seed photons are assumed to be produced at the shock front at a constant rate. 
Numerically, a fixed number of photons are injected at the shock front at every time step. 

For the velocity and density profiles around the shock front, we divide the spatial interval $[0,L]$ into $N_x$ equal zones with the width $\Delta x=L/N_x$. 
Thus the spatial coordinate of the center of the $i$-th zone is evaluated as
\begin{equation}
x_i=\Delta x(i-1/2)\ \ \mathrm{for}\ \ 1\leq i\leq N_x.
\end{equation}
We can derive these profiles at any time from the self-similar solution in \S2.1 evaluated at each of $x_i$.
The interval must cover the region from the initial to the final position of the shock. 
In the following calculation, we use the mesh covering the space $[0,2X_\mathrm{i}]$ ($L=2X_\mathrm{i}$) with 1000 zones ($N_x=1000$). 
The total number of seed photons is about $10^5$. 

\subsubsection{Photon propagation}\label{propagation}
We calculate the path through which a photon propagates during a time interval $\Delta t=t^{n+1}-t^n$ using an optical depth $\tau^n$ defined as
\begin{equation}
\tau^n\equiv \sigma_\mathrm{T} c\Delta t\tilde{n}^n_\mathrm{e}.
\end{equation}
Here we approximate the region around the photon as a uniform shell with the electron number density $\tilde{n}^n_\mathrm{e}$. 
This density is evaluated at the position $x$ where the photon is located at the time $t=t^n$ as
\begin{equation}
\tilde{n}^n_\mathrm{e}=\frac{(x-x_i)\tilde{n}^n_{\mathrm{e},i+1}+(x_{i+1}-x)\tilde{n}^n_{\mathrm{e},i}}{x_{i+1}-x_i}.
\end{equation}
Here the photon is supposed to be located in a position between $x_i$ and $x_{i+1}$.

The photon may be scattered by an electron within this time interval. 
The probability $P$ of this event is expressed by the optical depth as
\begin{equation}
P=1-\exp(-\tau^n).
\end{equation}
Generating a random number $R_1$ ranging from zero to unity, it is determined whether the photon travels freely or is scattered by an electron. 
When the random number $R_1$ is larger than the probability $P$, the photon travels freely. 
On the other hand, when $R_1$ is smaller than $P$, then the photon travels by a distance of 
\begin{equation}
l=-\frac{\ln(1-R_1)}{\sigma_\mathrm{T} \tilde{n}},
\end{equation}
before the scattering. 
In this case, we carry out the procedure for the scattering given in \S\ref{scattering}. 
After that, we evaluate the optical depth from the residual distance $c\Delta t-l$ as $\tau=\sigma_\mathrm{T}\tilde{n}^n_\mathrm{e}(c\Delta t-l)$ and then repeat the procedure above.

\subsubsection{Compton scattering\label{scattering}}
Since there is a well developed Monte-Carlo technique to deal with Compton scattering of  photons  \citep[e.g.][]{pss77,pss83}, we adopted this method and review the outline in this section.

At first, we specify the momentum $(p_x,p_y,p_z)$ of an electron. 
In our model, the momentum distribution $f_\mathrm{e}(p_x,p_y,p_z)$ of electrons at the scattering site is assumed to be a shifted Maxwell-Boltzmann distribution with the temperature $T_\mathrm{e}$, 
\begin{equation}
f_\mathrm{e}(p_x,p_y,p_z)\propto \exp\left[-\frac{(p_x-m_\mathrm{e}U_\mathrm{e})^2+p_y^2+p_z^2}{2m_\mathrm{e}kT_\mathrm{e}}\right],
\end{equation}
where $m_\mathrm{e}$ is the electron mass. 
In the following, the electron temperature $T_e$ is assumed to be the same as the photon temperature, i.e., $kT_\mathrm{e}=kT_\mathrm{ph}=0.1\,\mathrm{keV}$. 
The other parameter characterizing the above distribution is the bulk velocity $U_\mathrm{e}$ of electrons at the scattering site.  
In the same way as we calculate the electron number density in \S\ref{propagation}, a linear interpolation of the velocity profile is used to calculate the value of $U_\mathrm{e}$. 
For given $T_\mathrm{e}$ and $U_\mathrm{e}$, the momentum of the electron is determined using random numbers. 

We express the 4-momentum of the photon before and after scattering in the rest frame of the electron as 
\begin{equation}
{\bf p}_\mathrm{i}=\left(\begin{array}{c}
\epsilon_\mathrm{i}/c\\\epsilon_\mathrm{i}\sin\theta_\mathrm{i}\cos\phi_\mathrm{i}\\\epsilon_\mathrm{i}\sin\theta_\mathrm{i}\sin\phi_\mathrm{i}\\\epsilon_\mathrm{i}\cos\theta_\mathrm{i}
\end{array}\right)
,\ \ \ 
{\bf p}_\mathrm{f}=
\left(\begin{array}{c}
\epsilon_\mathrm{f}/c\\\epsilon_\mathrm{f}\sin\theta_\mathrm{f}\cos\phi_\mathrm{f}\\\epsilon_\mathrm{f}\sin\theta_\mathrm{f}\sin\phi_\mathrm{f}\\\epsilon_\mathrm{f}\cos\theta_\mathrm{f}
\end{array}\right),
\end{equation}
respectively. 
We use the Thomson cross section rather than the Klein-Nishina cross section for Compton scattering, because energies of photons considered here are significantly lower than the electron rest energy. 
Thus the differential cross section $d\sigma/d\Omega$ is given by
\begin{equation}
\frac{d\sigma}{d\Omega}=\frac{3}{16\pi}\sigma_\mathrm{T}(1+\cos^2\Theta), 
\label{dsigma}
\end{equation}
\citep[see e.g.,][]{rl85}.
Here we introduce the angle $\Theta$ between the directions of a photon before and after scattering,
\begin{equation}
\cos\Theta=\cos\theta_\mathrm{i}\cos\theta_\mathrm{f}+\sin\theta_\mathrm{i}\sin\theta_\mathrm{f}\cos
(\phi_\mathrm{i}-\phi_\mathrm{f}).
\end{equation}
We determine the angles  $\theta_\mathrm{f}$ and $\phi_\mathrm{f}$ after scattering by two random numbers so that the probability of the photon scattered into each solid angle follows  the above differential cross section (\ref{dsigma}). 
We obtain the energy of the scattered photon from the well-known formula \citep[e.g.,][]{rl85},
\begin{equation}
\frac{\epsilon_\mathrm{f}}{\epsilon_\mathrm{i}}=\frac{m_\mathrm{e}c^2}{m_\mathrm{e}c^2+\epsilon_\mathrm{i}(1-\cos\Theta)}.
\end{equation}

After we obtain the 4-momentum of the photon after scattering, we apply the inverse Lorentz transformation to the 4-momentum ${\bf p}_\mathrm{f}$ to obtain that in the original frame. 

\subsection{Derivation of light curves\label{lc}}

\begin{figure}
\begin{center}
\includegraphics[scale=0.5]{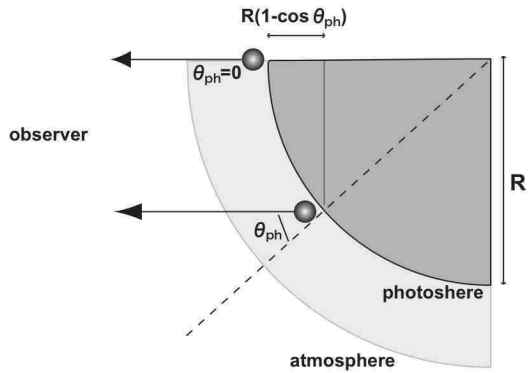}
\caption{A schematic view of the geometry of the emission region of X-ray photons. }
\label{f3}
\end{center}
\end{figure}

Since the adopted self-similar solution assumes the plane-parallel atmosphere and we only consider a region near the surface, we need to take into account several effects when deriving light curves of photons produced by the shock wave. 
We describe the procedure to derive light curves in this section. 

After we stop the calculation of the propagation of photons around the shock front, we evaluate the optical depth from the location of the photon to the surface $\tau_\mathrm{ph}$. 
First of all, we assume that non-thermal photons inside the photosphere are thermalized before they escape from the stellar surface. 
Thus, we only deal with photons whose optical depth is smaller than unity, $\tau_\mathrm{ph}<1$, to derive light curves of non-thermal emission. 
Next, we calculate the propagation of the thus selected photons toward the surface according to the procedure described in \S 2.3. 
Then, for each photon, we obtain the angle $\theta_\mathrm{ph}$ between the direction of the propagation of the photon and the shock normal when the photon reaches the surface. 
Finally, we consider light travel effects by using the angle $\theta_\mathrm{ph}$. 
The situation is schematically shown in Figure \ref{f3}. 
We assume a spherical explosion. 
As a consequence, the shock breakout occurs simultaneously and in the same way at every points on the surface of the progenitor star. 
Then, we assume that a photon with $\theta_\mathrm{ph}$ is emitted at a point  on the surface of the star, where the angle between the line of sight and the radial direction is equal to $\theta_\mathrm{ph}$. 
The time $t$ is measured from the point when photons with $\theta_\mathrm{ph}=0$ are observed. Then photons with $\theta_\mathrm{ph}$ are observed at $t=R(1-\cos\theta_\mathrm{ph})/c$ due to a difference of the path from the emitting point (see Figure \ref{f3}), where $R$ is the radius of the photosphere of the star. 
Taking the above two effects, i.e., the optical depth and the light travel effect, into account, we derive light curves of X-ray photons emitted after supernova shock breakout.

\section{RESULTS}
We calculate spectra and light curves of photons having experienced the bulk comptonization after supernova shock breakout according to the procedure described in the previous section. 
In this section, we apply our model to the observed X-ray spectrum and light curve of XRF 080109/SN 2008D. 

The progenitors of SNe Ib and Ic are expected to be Hydrogen-deficit compact stars, whose typical radius is of the order of $10^{11}$ cm. 
On the other hand, the long duration of XRF 080109, $\Delta t\sim 600$ s, predicts a much larger scale of the emitting region, i.e., $c\Delta t\sim 2\times 10^{13}$ cm. However, as previous works \citep{c06,w07,wwm07} have claimed, we can attribute the contradiction to the presence of a dense stellar wind. 
Namely, the shock breakout occurs in the dense wind that has the photosphere with a much larger radius than the stellar envelope. 
Then, we adopt $R=10^{13}$ cm as the radius of the photosphere. 
This modification does not alter the conclusion in \S \ref{radiative_process}, i.e., we can neglect photoionization, radiative recombination, and free-free absorption. 

\subsection{Spectra}
\begin{figure}
\begin{center}
\includegraphics[scale=0.5]{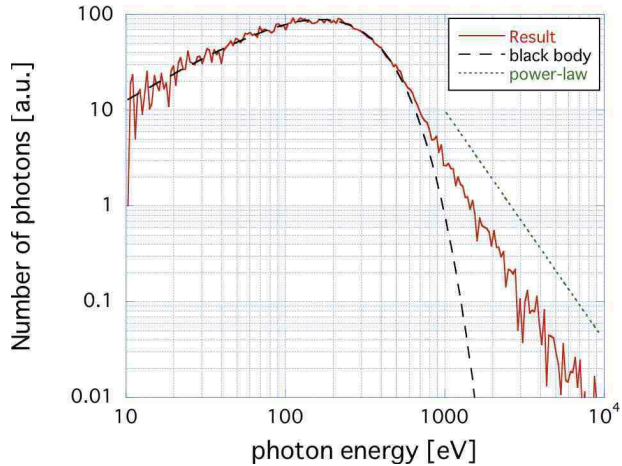}
\caption{An example of the resultant energy spectra of photons experienced the bulk comptonization process in supernova shock breakout (solid line). 
The free parameters characterizing the shock profile are $V_\mathrm{i}=0.3c$ and $X_\mathrm{i}=10^{11}\mathrm{cm}$. 
The radiative envelope of a massive star ($\delta =3$) is assumed. 
A black body spectrum with a photon temperature of $0.1$ keV (dashed line) and a power-law spectrum with the exponent of $-2.4$ (dotted line) are also plotted.}
\label{f4}
\end{center}
\end{figure}

Figure \ref{f4} shows an example of energy spectra of photons calculated from a shock breakout model presented in Figure \ref{f1}. 
To obtain the spectra, we use photons outside the photosphere ($\tau=1$). 
The values of the free parameters taken for this model are $V_\mathrm{i}=0.3c$ and $X_\mathrm{i}=10^{11}$ cm. 
The dashed line represents a black body spectrum with the photon temperature of $kT_\mathrm{ph}=0.1$ keV. 
While the resultant spectrum is consistent with the black body spectrum in a low energy range ($10$ eV-1 keV), it significantly deviates from the black body above $1$ keV. 
The dotted line represents a power-law spectrum with the exponent of $-2.4$, which is implied by the observed spectrum of XRF 080109 \citep{s08,l08}. 
The resultant spectrum above $1$ keV remarkably reproduces the observed power-law. 

In order to check the validity of the calculation, we have calculated the energy spectrum of photons injected at a point which is in the downstream and is sufficiently far from the shock front.
We confirmed that the spectrum remains black body one with $T_\mathrm{ph}=0.1$ keV. 
Therefore, we can attribute the formation of the power-law tail in the energy spectrum to the interaction of photons with the shock front. 

\begin{figure}
\begin{center}
\includegraphics[scale=0.5]{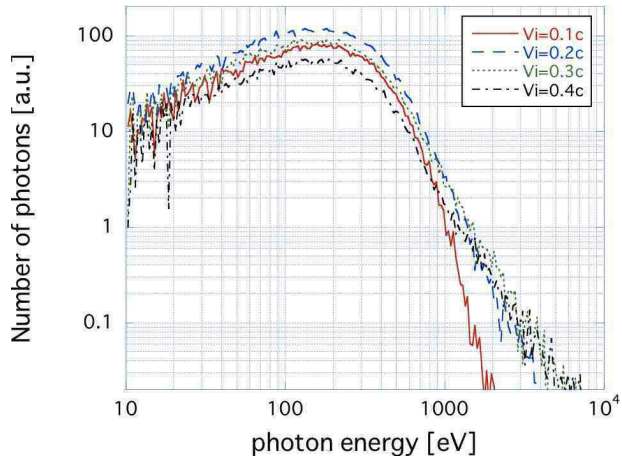}
\caption{The energy spectra for various initial velocity $V_\mathrm{i}$. 
Each line represents the model with $V_\mathrm{i}=0.1c$ (solid), $V_\mathrm{i}=0.2c$ (dashed), $V_\mathrm{i}=0.3c$ (dotted), and $V_\mathrm{i}=0.4c$ (dash-dotted). 
The initial position of the shock is fixed ($X_\mathrm{i}=10^{11}$ cm).
The radiative envelope of a massive star ($\delta =3$) is assumed. }
\label{f5}
\end{center}
\end{figure}

Figure \ref{f5} shows energy spectra of photons calculated for various $V_\mathrm{i}$ and fixed $X_\mathrm{i}(=10^{11}\,\mathrm{cm})$. 
Four different lines represent models with $V_\mathrm{i}=0.1c$ (solid), $V_\mathrm{i}=0.2c$ (dashed), $V_\mathrm{i}=0.3c$ (dotted), and $V_\mathrm{i}=0.4c$ (dash-dotted). 
We can easily recognize a tendency in the spectra that a larger shock velocity model has a more prominent high energy tail deviating from the black body spectrum. While the power-law tail is absent in the spectrum of models with $V_\mathrm{i}<0.1c$, 
the exponent of the power-law part of the spectra converges to $\Gamma \sim -2$ for models with the initial shock velocity larger than $0.4c$. 
However, since the shock profile we used here are valid for non-relativistic shock velocity, the value of the photon index may change for relativistic cases.

\begin{figure}
\begin{center}
\includegraphics[scale=0.5]{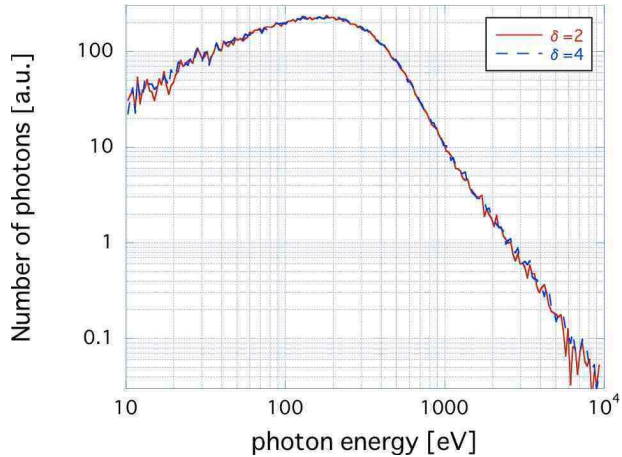}
\caption{The energy spectra for various $\delta$.  
Each line represents the model with $\delta =2$ (solid) and $\delta=4$ (dashed). 
The initial velocity and the initial position of the shock are fixed ($V_\mathrm{i}=0.3c$ and $X_\mathrm{i}=10^{11}$ cm).}
\label{f6}
\end{center}
\end{figure}

Finally, we investigate the dependence on the density profile. 
The density profile of the stellar envelope is characterized by the parameter $\delta$ in our model. 
We performed calculation of models with $\delta=2$ and $\delta=4$. 
The resultant energy spectra are shown in Figure \ref{f6}. 
We found that these energy spectra also show power-law tails with the same index as the case for $\delta=3$, which is a physically natural consequence. 
High energy photons comprising the power-law tail are produced by repeated crossing of the shock front \citep{bp81a,bp81b}. 
The probability that photons diffusing out from the shock front scattered back to the downstream is determined by the optical depth of the envelope, which is given by integrating the density profile, rather than the density profile itself. 
In addition, as long as the optical depth is larger than unity, most of photons must be scattered back to the downstream. 
Slight differences of the density structure do not alter the bulk comptonization process. 
Therefore, the results weakly depend on the parameter $\delta$. 

\subsection{Light curves}
\begin{figure}
\begin{center}
\includegraphics[scale=0.5]{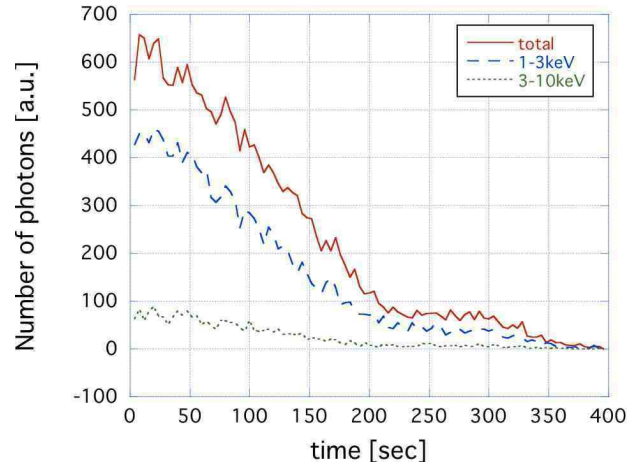}
\caption{An example of light curves of X-ray photons ($1$ keV$<h\nu<10$ keV) calculated using the same model as in Fig. \ref{f4}. 
Different lines represent the numbers of photons in the energy ranges of $1$-$10$ keV (solid), $1$-$3$ keV (dashed), and $3$-$10$ keV (dotted) as functions of time.}
\label{f7}
\end{center}
\end{figure}

Using the procedure described in \S\ref{lc}, we calculate light curves of X-ray photons produced at the shock front. 
One of the results is shown in Figure \ref{f7}. 
The radius of the progenitor star is assumed to be $10^{13}$ cm, which is required to explain the observed duration of the outburst of a few hundred seconds. 

We mention an inadequacy of our light curve model. 
Results of optical spectroscopic observations of the nebular phase of SN 2008D \citep{t09} strongly indicate the aspherical nature of SN 2008D. 
On the contrary, our model assumes a spherical explosion, which leads to a consequence that the shock breakout simultaneously occurs at every point on the surface of the progenitor star. 
For an aspherical explosion, the shock breakout does not occur simultaneously in different radial directions. 
The assumption of spherical symmetry is not appropriate when the duration of the shock breakout is comparable to that of the X-ray outburst. 
This effect modifies results of our light curve model. 
However, the investigation into this asynchronism requires two-dimensional hydrodynamical simulations of supernova explosions. 
Therefore, we consider that the modeling of light curves of X-ray photons from an aspherical explosion is beyond the scope of this work. 
Furthermore, the light curve of X-ray emission from XRF 080109 have large errors due to the paucity of photons. 
Therefore, it is difficult to judge whether SN 2008D was a spherical or aspherical explosion using only our results assuming plane-parallel atmosphere. 

\subsection{Energetics}
Finally, we check the consistency of our model with the observational results. 
Since we have assumed the black body radiation with the photon temperature of $T_\mathrm{ph}=0.1$ keV and the scale of the emitting region to be $R=10^{13}$ cm, the luminosity $L_\mathrm{th}$ of the emission is of the order of $10^{47}$ erg/s. 
The time during which the shock wave emits thermal photons, $(X_\mathrm{i}-X_\mathrm{f})/V_\mathrm{s}$ is of the order of seconds. 
Therefore, the total energy of the thermal emission is evaluated as $E_\mathrm{th}=10^{47}$ erg. 
On the other hand, from the results of the simulation (Fig. \ref{f4}), the ratio of the energy of non-thermal X-ray photons with 0.3-10 keV to that of the thermal emission is evaluated as $\epsilon=0.3$-$0.4$. 
Using these values, we obtain the total emitted energy of non-thermal X-ray photons with 0.3-10 keV, $E_\mathrm{X}=\epsilon E_\mathrm{th}=$(3-4)$\times10^{46}$ erg, which agrees with the observed value very well. 
In other words, adopting the progenitor radius of $10^{13}$ cm, we can explain both the duration and the total energy of the observed X-ray emission. 
It should be noted that the above estimate has an uncertainty. 
To obtain the spectrum shown in Fig. \ref{f4}, we use photons outside the photosphere ($\tau=1$). 
However, photons in the downstream of the shock must diffuse out later and contribute to the thermal part of the spectrum. 
Especially, photons in the region where $\tau<c/V_\mathrm{s}$ can escape from the shock front. 
Therefore, we may overestimate the ratio $\epsilon$ by a factor of $c/V_\mathrm{s}$. 

\section{CONCLUSION}
In this paper, we calculate the propagation of photons at the supernova shock breakout by using a Monte-Carlo technique in order to explain the power-law spectrum revealed by X-ray observations for XRF 080109/SN 2008D in the framework of the shock breakout origin of the X-ray photons. 

We use the self-similar solution for the propagation of a shock wave in a stellar envelope presented in \cite{s60} to acquire the velocity profile and evaluate the optical depth for photons around the shock front. 
Our hydrodynamical model is more realistic than that used in \cite{wwm07} and allows us to obtain plausible constraints on the nature of the shock breakout occurred in SN 2008D. 
Our simulations clearly demonstrate that a mildly relativistic shock velocity, $V_\mathrm{s}>0.3c$, is required for the spectrum of photons emitted from the shock front to deviate from the black body spectrum due to repeated scatterings around the shock, i.e., the bulk comptonization. 
The resultant spectra successfully reproduce the power-law spectrum with the exponent of $\sim 2.4$ seen in the X-ray spectrum of XRF 080109. 
We also investigate whether some radiative processes, photoionization, radiative recombination, and free-free absorption, interrupt the formation of the power-law tail and find that none of them hardly changes the shape of the spectra as long as progenitor stars of type Ib or Ic supernovae are concerned. 

Assuming a spherical explosion, light curves of the X-ray photons are derived. 
To account for the duration of the X-ray emission from XRF 080109, the progenitor radius of $\sim 10^{13}$cm is needed. 
Although this value is somewhat large compared to the typical radius of Wolf-Rayet stars, the shock breakout taking place in a dense stellar wind may explain the discrepancy. 
Furthermore, our model successfully explain the total energy of the observed X-ray emission. 
Because the density structure of the wind is not known, it is unclear whether our power-law approximation of the atmosphere is appropriate for the situation. 
However, the fact that our results weakly depend on the parameter $\delta$ suggests that the bulk comptonization can operate even in the dense wind and produce a power-law X-ray spectrum. 

Our model does not include the contribution of photons from the downstream of the shock and the matter ejected at $t>0$. 
However, as we noted in \S 3.1, photons originating from far downstream have a black body spectrum. 
So they do not contribute to non-thermal photons. 
In the same way, photons emitted by the ejected matter are not important. 
In particular, after the shock breakout, the ejected matter cools rapidly due to adiabatic expansion. 
It must emit optical or UV photons rather than X-ray photons. 
Therefore, both of them hardly change at least the high energy part of the energy spectra calculated using our model. 
As we noted above, however, these thermal photons may decrease the ratio of thermal to non-thermal emission from the value obtained in \S 3.3.

It should be noted that our model to calculate the radiation from the supernova shock breakout may have  a few inadequacies. 
The used self-similar solution cannot deal with highly relativistic flows and the photon temperature is assumed to be equal to the gas temperature even after the shock breakout occurs. 
Such caveats are improved by calculating the radiation from the supernova shock breakout by using more sophisticated simulation codes, e.g, a multi-group radiation hydrodynamics code.

\acknowledgments
We are grateful to M. Tanaka for making useful comments. 
We are grateful to the anonymous referee for his/her constructive comments on this manuscript. 
This work has been partly supported by  Grant-in-Aid for JSPS Fellows (21$\cdot$1726) and Scientific Research on Priority Areas (21018004) of the
Ministry of Education, Culture, Sports, Science, and Technology in
Japan.

\appendix
\section{DERIVATION OF THE SELF-SIMILAR SOLUTION}
In this section, we review the procedure to determine the eigen value $\lambda$. 
The following steps are originally presented in \cite{s60}. 

At first, we solve Eq. (\ref{dhdx}) with respect to $dg/d\xi$,
\begin{equation}
\frac{dg}{d\xi}=\frac{g}{1-\xi f}\left[\gamma\xi\frac{df}{d\xi}-\frac{\delta-(\gamma+2)\lambda}{1+\lambda}f\right].
\end{equation}
Substituting the above expression into Eq. (\ref{dgdx}), the following equation is obtained,
\begin{equation}
(1-\xi f)\frac{df}{d\xi}-\frac{\xi g}{h(1-\xi f)}\left[\gamma\xi\frac{df}{d\xi}-\frac{\delta-(\gamma+2)\lambda}{1+\lambda}f\right]
=\frac{\lambda}{1+\lambda}f^2-\frac{\delta-2\lambda}{1+\lambda}\frac{g}{h}.
\label{eq4}
\end{equation}
Next, subtraction of both sides of Eq. (\ref{dfdx}) from those of Eq. (\ref{dgdx}) leads to the following equation,
\begin{equation}
(1-\xi f)\left(\frac{d\ln g}{d\xi}-\frac{d\ln h}{d\xi}\right)=(\gamma-1)\xi \frac{df}{d\xi}+\frac{\gamma+1}{1+\lambda}\lambda f.
\label{eq5}
\end{equation}
Here we introduce new variables $y$ and $z$ as functions of $\xi$,
\begin{equation}
f=\frac{1}{\xi}\left(1-\frac{1}{z}\right),\ \ \ \frac{g}{h}=\frac{1}{\gamma\xi^2}\frac{y}{z^2}.
\label{define}
\end{equation}
Using these functions, Eqs. (\ref{eq4}) and (\ref{eq5}) are transformed into,
\begin{equation}
(1-y)\frac{d\ln z}{d\ln\xi}=(1-y)(z-1)-\frac{\delta-(\gamma+2)\lambda}{(1+\lambda)\gamma}y(z-1)
+\frac{\lambda}{1+\lambda}(z-1)^2-\frac{\alpha-2\lambda}{(1+\lambda)\gamma}y,
\label{eq4'}
\end{equation}
and
\begin{equation}
\frac{d\ln y}{d\ln \xi}=(\gamma+1)\frac{d\ln z}{d\ln\xi}+\frac{2\lambda-(\gamma-1)}{1+\lambda}z+\frac{\gamma+1}{1+\lambda}.
\end{equation}
Then, the elimination of the variable $\xi$ from the above two expressions yields
\begin{equation}
y\frac{dz}{dy}=\frac{\gamma(z-1)(\lambda z+1)+\left[(2\lambda-\delta-\gamma)z+\gamma\right]y}
{2\gamma-\gamma(\gamma-1)\lambda-[\delta +(\delta +2)\gamma-2\lambda]y+\gamma(\gamma+1)\lambda z}.
\label{dzdy}
\end{equation}
We solve this equation numerically to obtain the eigen value $\lambda$. 

Here we consider the boundary conditions of the equation. 
From the definition of the variables (\ref{define}) and the boundary condition at the shock front (\ref{bc}), their values $y_\mathrm{f}$ and $z_\mathrm{f}$ at the front are obtained as 
\begin{equation}
y_\mathrm{f}=\frac{2\gamma}{\gamma-1},\ \ \ 
z_\mathrm{f}=\frac{\gamma+1}{\gamma-1}
\end{equation}
On the other hand, another boundary condition is obtained as follows. 
In fact, Eq. (\ref{dzdy}) have a singular point. 
From Eq. (\ref{eq4'}), we can easily find that the singular point is located at $y_\mathrm{s}=1$, where the L.H.S of Eq. (\ref{eq4'}) vanishes. 
Then, a requirement that the R.H.S of Eq. (\ref{eq4'}) also vanishes at $y_\mathrm{s}=1$ leads to the value of $z$ at the singular point, 
\begin{equation}
z_\mathrm{s}=\frac{\delta-(2-\gamma)\lambda}{\gamma\lambda}. 
\end{equation}
From the thus derived boundary values, we can determine the eigen value $\lambda$ so that Eq. (\ref{dzdy}) can be integrated from $(y_\mathrm{f},z_\mathrm{f})$ to $(y_\mathrm{s},z_\mathrm{s})$. 
For $\delta=3$ and $\gamma=4/3$, which are used in this work, the eigen value is found to be $\lambda=0.55727$. 
The values of $\lambda$ for various $\delta$ and $\gamma$ are found in the original work \citep{s60} or \cite{ss07}. 
We obtain the profiles $f$, $g$, and $h$ by numerically integrating Eqs. (\ref{dfdx})-(\ref{dhdx}) for the eigen value.


\end{document}